\documentclass[12pt,preprint]{aastex}
\usepackage{epsfig}
\usepackage{amsmath}
\usepackage{graphicx}

\makeatletter

\usepackage{fancyhdr}

\makeatother
\shorttitle{Composition of Super-Earths}
\shortauthors{Valencia et al.}
\begin{document}

\title{Detailed Models of super-Earths: How well can we infer bulk properties?}

\author{Diana Valencia\altaffilmark{1}}
\affil{Earth and Planetary Science Dept., Harvard University, 20 Oxford Street, Cambridge, MA, 02138}
\email{valencia@mail.geophysics.harvard.edu}

\author{Dimitar D. Sasselov}
\affil{Harvard-Smithsonian Center for Astrophysics,
60 Garden Street, Cambridge, MA, 02138}

\and
 
\author{Richard J. O'Connell}
\affil{Earth and Planetary Science Dept., Harvard University, 20 Oxford Street, Cambridge, MA, 02138}

\altaffiltext{1}{corresponding author}

\begin{abstract}
The field of extrasolar planets has rapidly expanded to include the
detection of planets with masses smaller than that of Uranus. Many
of these are expected to have little or no hydrogen and helium gas
and we might find Earth analogs among them. In this paper we describe
our detailed interior models for a rich variety of such massive terrestrial
and ocean planets in the 1-to-10 $M_{\oplus}$ range (super-Earths). The grid
presented here allows the characterization of the bulk composition
of super-Earths detected in transit and with a measured mass. We show
that, on average, planet radius measurements to better than 5\%, combined
with mass measurements to better than 10\% would permit us to distinguish
between an icy or rocky composition. This is due to the fact that
there is a maximum radius a rocky terrestrial planet may achieve for
a given mass. Any value of the radius above this maximum terrestrial
radius implies that the planet contains a large ($>10$\%) amount of
water (ocean planet).
\end{abstract}


\section{Introduction}

A new family of exo-planets is now being explored thanks to improvements
in detection methods that allow the discovery of very low mass objects
(1-10 $M_{\oplus}$ hereinafter super-Earths). The first three super-Earths
discovered are: GJ876d - a 7.5 $M_{\oplus}$ planet at 0.02 AU from
its star \citep{Rivera_et_al:2005}, OGLE-2005-BLG-390Lb - a 5 $M_{\oplus}$
planet at 5 AU \citep{OGLE-5.5:2006} and HD69830b - a 10.8$M_{\oplus}$
at 0.08 AU \citep{Lovis_et_al:2006}. In the next few years, with
missions like COROT, that launched in December 2006 (\citet{COROT:2003})
and \emph{Kepler}, scheduled to launch in 2008 \citep{Kepler-Mission}, many
more - and smaller, super-Earths are expected to be found. These missions
detect planets by observing their transits, hence allow for a measurement
of the planet's radius, $R$.  The Doppler techniques have improved
significantly and allow detection and planet mass, $M$, determination
for super-Earths as well, as demonstrated by the discovery of GJ876d
and HD69830b.

Our work is motivated by the need for mass-radius relations for super-Earths
of varied compositions. Such relations should allow discerning the
bulk compositions of planets detected in transit - with derived $R$
and $M$. Zero-temperature sphere models \citep{Zapolsky_Salpeter:1969, Stevenson:1982-Giant-Planets, Fortney:2006} allow approximate mass-radius calculations for specific
simple materials (pure Fe, pure olivine, pure water). \citet{Valencia_et_al:2006}
introduced detailed interior models for super-Earths. These models
were extended to super-Earths with large water content \citep{Valencia_GJ876d}
- the ocean planets proposed by \citet{kuchner:2003,Leger_et_al:2004}.
A cold ocean planet model was introduced by \citet{Ehrenreich_aph:2007}
for the low-temperature planet OGLE-2005-BLG-390Lb. 

Distinguishing between ocean planets and dry rocky super-Earth planets
is not trivial, because at a given mass the same planet radius could
correspond to different mixtures of iron core, mantle minerals, and
ices \citep{Valencia_GJ876d}. Observations with certain tolerances
of the derived planetary $R$ and $M$ are required, as shown for
two particular cases by \citet{Selsis_aph:2007}. On the other hand, there
is very strong motivation to be able to distinguish between ocean
planets and dry planets, because of all the implications to planet
formation theory and the development of habitable environments. Models
of planet formation \citep{Ida:2004, Rafikov:2006,
Kennedy_Kenyon:2006,Raymond:2006} anticipate
the formation of super-Earths with a rich variety of bulk compositions,
but differ in the details. 

As the field of planets with masses in the range of 1-10$M_{\oplus}$
emerges, it will be useful to have a consistent nomenclature to describe such planets. We propose to call a super-Earth,
a planet with large amounts of solid material in the 1-10$M_{\oplus}$
range. The lower bound is obvious for historical reasons, the upper bound is
somewhat arbitrary.  It is informed by the physical argument that at
$\sim 10M_\oplus$ and above, the planet can retain H and He during formation \citep{Ida:2004}.
Super-Earths, includes ocean planets and massive
terrestrial planets as a subclass in this mass range. Planets with H$_{2}$O on the order of $\geq10$\%
in either liquid and/or solid phase are termed ocean planets.
Massive terrestrial planets will then be the ones with little water
on the surface or in the mineral assemblages and hence, be the most
similar to Earth. The existence of a large core of iron makes a difference,
therefore within super-Earths there is a subset of planets
with large cores that we have termed 'super-Mercuries', and which
can be terrestrial or have a large ocean. The name 'ice giants' -
Uranus and Neptune - has been reserved for planets more massive than
$10M_{\oplus}$ with some amount of gas ($\sim10$\%). In addition,
it is possible, though unlikely, that planets with $M<10M_{\oplus}$
could have a lot of gas ($>10$\%), but we do not model them in this
paper. At $10M_{\oplus}$, a complication arises due to the possibility
of H and He retention by the planet during formation. We do not consider
these planets in this paper as a separate class since they can be
treated as massive ocean planets with high H$_{2}$O content and a
relatively small layer of gas above it that would add a few kilometers
to the radius. The regime of $M<1M_{\oplus}$, that includes Mars,
Venus, Mercury and the icy-satellites of the Solar System has been
studied previously with models describing these planets and therefore
has been left out of this paper. 

In this study we explain how we can differentiate between a rocky
and an ocean super-Earth with measurements of mass ($M$) and radius
($R$) at the current precision levels. In the first section we briefly
explain the model used to obtain the internal structure of a super-Earth
given its composition ($\chi$) and $M$. This model has been used
to determine the total radius and interior distribution of bulk parameters
(mass, gravity, pressure and temperature) of terrestrial and ocean
exo-planets \citep{Valencia_et_al:2006,Valencia_GJ876d} and here
we present another useful application of this model: how to obtain
constraints on $\chi$ with values of $M$ and $R$.

\section{Model}

We model the internal structure of a low mass exo-planet by dividing
its interior in concentric shells where composition is considered
homogenous. Within each layer the numerical model solves the standard
differential equations for density, pressure, mass and gravity structure
under hydrostatic equilibrium. The equation of state (EOS) chosen
to relate density, pressure and temperature is the Vinet EOS \citep{Vinet:1989}
that extrapolates better to high pressures than the third-order Birch-Murnaghan
(BM3) EOS \citep{Hama_Suito:1996, Cohen_Gulseren_Hemley:2000}. The
regions are:

\begin{itemize}
\item an H$_{\textrm{2}}$O layer divided into a water ocean over an icy
shell - composed of high-pressure ices VII and X 
\item the mantle divided into an upper mantle - composed of olivine ({[}Mg$_{1-x}$,Fe$_{x}$]$_{2}$SiO$_{4}$)$^{\alpha}$,
wadsleyite ({[}Mg$_{1-x}$,Fe$_{x}$]$_{2}$SiO$_{4}$)$^{\beta}$
and ringwoodite ({[}Mg$_{1-x}$,Fe$_{x}$]$_{2}$SiO$_{4}$)$^{\gamma}$
- and a lower mantle - composed of perovskite ({[}Mg$_{1-x}$,Fe$_{x}$]SiO$_{3}$),
ferromagnesiowusite ({[}Mg$_{1-x}$,Fe$_{x}$]O) and post-perovskite
({[}Mg$_{1-x}$,Fe$_{x}$]SiO$_{3}$)
\item and the core (mainly Fe), that can be divided into a liquid outer
layer and inner solid layer if the temperature profile of the planet
crosses the melting curve of iron. 
\end{itemize}
The thickness of the H$_{2}$O, mantle and core regions are dependent
on the ice mass fraction (IMF) and core mass fraction (CMF) that are
assumed for a planet. The Earth has a negligible water fraction (IMF=0.02-0.1\%)
and a core that is 1/3 of its mass (CMF=32.59\%, \citet{AppendixF-Stacey:Physics-Earth}).
Phase transitions are modeled in each region to obtain the different
layers composing the planet. The temperature profile is modeled after
the Earth's; the temperature gradient is adiabatic within each region
and conductive within the boundary layers developed at the top and
bottom of the mantle. The boundary layers in the H$_{2}$O and core
regions are neglected owing to their very low viscosity that prevents
thick boundary layers to develop. The surface temperature is taken to be $500$
K after the detection of GJ876d with an estimated
surface temperature of $430-650$K  \citep{Rivera_et_al:2005}.
Despite this high value, water may be present provided the surface pressure
is large enough ($\sim 1000$ atm for $T_{surf}=500$). For more details please refer to
\citet{Valencia_et_al:2006,Valencia_GJ876d} where the model is explained
thoroughly. Currently the models are not evolved in time; they correspond to a
planet that has undergone differentiation and achieved equilibrium like our
Earth. 

This numerical model allows us to calculate the total radius of a
planet given its mass and composition. We do not know the exact composition
of the planets that will be discovered in the near future but we can
make reasonable assumptions on their bulk composition. The dominant
minerals in the mantle are most likely the same as for Earth (silicates
with a small incorporation of iron -$\sim$10\%). The core is mainly
composed of Fe. The Earth's core is known to have some Ni (mostly
in the inner core) and a light alloy in the outer core of $\sim$10\%.
The nature of the light alloy and exact amount are not known for Earth. Instead of looking at all possible candidates (O, H, S, Si,
C) we model the core as pure Fe in this study and anticipate that
the presence of a light allow will increase the radius by a few hundred
kilometers. The water region is composed of ice VII and X. For low
temperature planets, there might also be an extra icy layer of Ice
I above the water ocean. This layer will be small due to the small
stability field of Ice I and hence we consider the results presented
in this study to be adequate even for low temperature icy planets
(such as Ganymede, Europa and Calisto massive analogs).  For planets that are
very close to its parent star, their surface temperature might be on the order
of $500-1500$K.  Ice VII is a high pressure form of H$_2$O that could be the
source of a liquid water layer, provided the surface temperature is high
enough according to the phase diagram of H$_2$O \citep{Wagner_Pruss:water_2002}.  The thermal structure has little effect
on the radius of a planet \citep{Valencia_et_al:2006} but imposes conditions
on the compositional phases, particularily of H$_2$O at the surface. If the
surface temperature is higher than the critical point ($T_{crit}=647$K and
$P_{crit}=218$ atm) the structure of an ocean planet will vary gradually from a
vapor phase to an ice VII phase at depth.

\section{Ternary Diagram }

We have computed a grid of interior models appropriate for ocean and
terrestrial planets with the entire range of CMF and IMF to determine
how $M$, $R$ and $\chi$ are related. Since the percentage of mass
in the mantle, ices and core has to add up to 1, we use a ternary
diagram to show the results. Ternary diagrams are commonly used in
Earth sciences. Data of a three-component system is plotted in a triangle
whose sides depict three axes. Each vertex means 100\% of a particular
component and data plotted on the opposite side means 0\%. Lines parallel
to a particular side will show various degrees of a component whose
maximum value is shown at the corresponding opposite vertex. We use
a fourth axis (color code or label) to show the value for the radius
for each composition. A 3-dimensional volume describes the complete
family of Mass-Radius relations for all compositional combinations;
a ternary diagram is a cross-section at a given mass. Figure 1
shows some examples for a 5$M_{\oplus}$ planet (like OGLE-2005-BLG-390Lb)
with three different compositions (A) CMF=10\% and IMF=50\% with a
radius of 12200km, (B) CMF=20\% and IMF=10\% with a radius of 10750km
, (C) CMF=30\% and no water with a radius of 9800km and (D) CMF=40\%
and IMF=30\% with a radius of 10950km.

A minimum and maximum radius exists for a planet with mass $M$. The
smallest a planet can be is when it is composed of pure iron or heavy
iron alloys (Fe+Ni) and is shown in the right vertex of the ternary
diagram. The maximum size for a semi-solid planet (not gaseous) is
obtained if it is composed purely of ices and water and plots in the
left corner. For a 1$M_{\oplus}$ planet these extreme values are
$\sim$4900 km and $\sim$9300 km. Simple formation constraints suggest
that the minimum and maximum are not quite at the vertices (more
realistic conditions see section 4.1). Terrestrial planets (no substantial
water content) of 5$M_{\oplus}$ are shown in Fig. 2 on the side connecting
mantle and core only. \\

\section{Results}

We have calculated the ternary diagrams for planets with 1$M_{\oplus}$,
2.5$M_{\oplus}$, 5$M_{\oplus}$, 7.5$M_{\oplus}$ and 10$M_{\oplus}$
showing the maximum and minimum radius for each planet and curves
of constant radius. Figure 2 shows the different ternary diagrams.
The same value for radius can be obtained through different combinations
of core, mantle and H$_{2}$O contents showing degeneracy in composition.

\subsection{Initial Conditions and Constraints on Bulk Compositions }

Our interior model of a super-Earth planet could not be a random mixture
of layers of different mineral, alloy, or compositional shells. This
is so for at least two general reasons: differentiation and constraints
in the proto-planetary disk abundances. Therefore we make two basic
assumptions in this work. The first is that all our model planets
have undergone differentiation. The second is that the relative elemental
abundances are the same as those in the Solar System and the solar
neighborhood. The first assumption is not restrictive since all the
terrestrial planets and large satellites in the Solar System are known
to be differentiated. The second assumption should also be adequate
for the super-Earths that we expect to be discovered in the near future
(close-by and in the pre-selected targets of the Kepler mission).

Thus, we consider three possible initial states for our models: 

(1) formation and post-differentiation anywhere in the proto-planetary
disk - the ratios of Si/Fe, O/Fe, etc. depend on $T_{cond}$ and the
amount of Fe incorporated in the mantle depends on the O locally available;

(2) formation as in (1), followed by a collisional stripping - then
four initial states are possible: (a) pure iron alloy core; (b) core
and mantle; (c) pure mantle; and (d) mantle and water ocean; and, 

(3) formation as in (1) or (2), followed by water delivery - then
two initial states are possible: (a) core, mantle, and water ocean,
and (b) mantle and water ocean.\\

To explore the first scenario we consider that during formation, the
building blocks of planets - volatiles, silicates and metals - condense
out at different temperatures $T_{cond}$. As the nebula cools down
the most refractory elements will condense out first. The first of
the building blocks of planets to condense out are silicates at temperatures
between 1750-1060K \citep{Pataev_Wood:2005}, followed by the metals
(Fe, Ni) between 1450-1050K \citep{Pataev_Wood:2005} and finally
H$_{2}$O and other ices, depending on the pressure. The relative
amounts of the different major elements in solid planets are H
at 74\%, O at 1.07\%, Fe at 0.1\%, Si at 0.065\% and Mg at 0.058\%.
We consider Si and Mg to be practically equally abundant. During the
condensation sequence (for pressure $<10^{-4}$bars), Si will condense
before Fe. If Fe remains immiscible it will form the core during condensation
and planet segregation. This is the end state scenario that yields
the largest core possible for a planet. That is, a fixed ratio of
Si/Fe of $\sim0.6$. The mantle is effectively MgO+SiO$_{2}$ so that
Si/Fe ratio can be used as a proxy for MMF/CMF. This line is shown
in Fig. 1 as the lowest possible
values for MMF/CMF. Any value above this line is plausible during
formation because Fe can be incorporated in the mantle to form (Mg$_{1-x}$,
Fe)O+SiO$_{2}$ (with $x=0.1$ for Earth) depending on the redox conditions.

To explore the possibility of having a planet with a large core and
H$_{2}$O layers and a small mantle we consider a scenario with a
pure Fe embrio that acquires water from comet delivery, which is scenario
(3) stated above. Comets have a dust/gas ratio of 1-2. As the comet
delivers water (gas, volatile) it will at least deliver Si+Mg+Fe+O
in the same proportions. We consider that the cometary dust is made
up of Si, Mg, Fe and O and that the proportions of these elements
are the same as in the solar nebula (or CI chondrites). Then the ratio
of Si/H$_{2}$O that needs to be maintained is at least 0.23 by mass.
Therefore, for every gram of water delivered to the Fe pure embrio,
0.23g of Si (that makes up the mantle) will be delivered also. This
is the end state case that yields the lowest amount of MMF possible
for a given IMF. This line shows the minimum amount of mantle formed
as planetesimals get accreted and deliver water. Accompanying this
amount of Si, is the amount of Fe that, if in solar abundances can
also add mass to the core. Therefore the unshaded region in Fig. 1
shows the plausible compositions that reflect the solar nebula composition
and cometary delivery constraints.

To explore scenario (2) mentioned above, we contemplate late-stage
processes that might influence the final accretionary state of a planet
starting from a 'normal' embrio. Any process that induces the escape
of preferentially light elements (solar wind, gravitational escape)
will deplete the planet from volatiles, H$_{2}$O and perhaps silicates
forming terrestrial planets (core, mantle+core planets). Conversely,
any process that targets heavy elements to be excluded from accretion
(like formation of the Moon after the Mars-sized impact onto a proto-Earth
\citep{Canup:2004}) would deplete the planet from Fe and silicates
forming a light planet (mantle, mantle+water ocean planets). It is
difficult to imagine a process that would preferentially retain the
heavy and light elements simultaneously and blow-off or prevent from
accreting the medium density components (silicates).

Based on these arguments we expect that planets with large CMF and
IMF and small or non-existent mantles are very unlikely. Thus, the
shaded region in Fig 1 is unlikely
to be populated.

\subsection{Maximum Radius for Terrestrial Planets}

For a given planet, a threshold in radius exists such that larger
values imply that the planet is ocean-like - has a substantial amount
of water and ices and can not be completely rocky (\citep{Valencia_GJ876d}.
This threshold radius can be identified as the largest iso-radius
curve that intersects the terrestrial side on Fig 2.
All the iso-curves to the left (bluer region) of this curve show IMF$>$0.
This threshold in radius is 6600 km, 8600km, 10400 km, 11600km and
12200km for 1$M_{\oplus}$, 2.5$M_{\oplus}$, 5$M_{\oplus}$, 7.5$M_{\oplus}$
and 10$M_{\oplus}$ planets respectively. Thus, for a given planetary mass,
a measurement in radius larger than its corresponding terrestrial
threshold radius would imply that the planet either formed beyond
the snow line acquiring large amounts of H$_{2}$O, or had large amounts
of volatiles delivered to it by wet planetesimals. 

The identification of the composition via this maximum terrestrial
radius will be easier for smaller planets since their iso-curves are
more significantly separated in the ternary diagram. \\

\subsection{Iso-radius curves}

Figure 2 shows the progression in radius from
a 10$M_{\oplus}$ to a 1$M_{\oplus}$ planet and how the iso-radius
curves move closer to each other as the mass increases. The lines
of constant radius are less sensitive to the amount of mantle mass,
illustrated in how parallel they lay with respect to increasing values
of mantle percentage (iso-curves almost vertical). 

\begin{itemize}
\item This is particularily true for a 5$M_{\oplus}$ planet: a large range
of mantle mass (between $0-85$\%) may yield the same radius of $\sim11300$km,
provided the correct amount of water is present (from 50\% to 15\%
respectively).
\item For a 1$M_{\oplus}$ planet, the iso-radius curves are tilted to be
slightly more sensitive to mantle mass, less sensitive to the amount
of core and more sensitive to water content (iso-cuves are more parallel
to the lines of constant water content). At this pressure regime (smallest
planet shown), the amount of low density (water) and high density
(Fe) material can proportionally affect the total radius. 
\item For the case of a 10$M_{\oplus}$ (high internal pressures) the iso-curves
show that for very water rich planets ( $>$ 70\% H$_{2}$O), the radius
is dependent on the amount of core-mass fraction (blue lines moving
towards being parallel to the terrestrial line) and become less independent
as the composition becomes iron rich ($>$ 70\% Core). For the large
core fractions, the amount of water is important because the contrast
in density is the largest. The overall trend is that the iso-radius
curves are less sensitive to the amount of mass in silicate form (mantle)
and the result for radius depends more in the amount of light elements
(water and ices) and heavy elements (core-Fe).
\end{itemize}
There have been useful mass-radius relationships obtained with
zero-temperature sphere models  \citep{Zapolsky_Salpeter:1969, Stevenson:1982-Giant-Planets, Fortney:2006} that hint to the bulk composition of a planet. These
models find the radii for different pure compositions
such as H$_{2}$O, silicates or Fe.  These radii (as a function of mass)
correspond to each vertex of our ternary diagram. They also mix pairs of the
above compositions, essentially following the sides of the ternary diagram. It can be seen from
Fig. 2 that the iso-radius curves are very sensitive to the mixture and how
they intersect with the ternary sides depends on it, especially in a
mantle-H$_{2}$O mixture. Of course, the zero-temperature sphere models have no
layering, or phase transitions, or convective mixing in the planet's
interior.  They also do not access the parameter space inside the ternary diagram.
Therefore, a zero-temperature model is a good first approximation. However,
detailed calculations are needed to distinguish between realistic compositions
at the precision CoRoT and \emph{Kepler} will allow.

\section{Mass-Radius Relationship}

We had previously stated that the radius of a super-Earth scales in
a power law relationship with mass: $R=aR_{\oplus}\left(\frac{M}{M_{\oplus}}\right)^{\beta}$
where $a=1$ and $\beta\sim0.27$. We have improved this result by
choosing an equation of state better suited for high-pressure extrapolations
(Vinet EOS) and including the recently discovered new silicate phase
- post-perovskite (ppv) in the Earth's lower-most mantle \citep{Motohiko:2004}.
Even though the stability field of ppv is only of $\sim10$GPa on
Earth, it constitutes most of the mantle of super-Earths. It also
can accomodate more Fe in its structure than perovskite \citep{Mao:2004}
making it more dense. With these improvements the exponent of the
power law becomes smaller: $\beta=0.262$. It is different from 1/3
- the constant density scaling value - due mostly to pressure effects.
The scaling is independent of temperature effects (such as surface
temperature, constant versus temperature-dependent viscosity cases),
and CMF. It also differs little with core composition.

Sotin et. al \citep{Sotin:2007} has shown that planets that have 50\%
water content scale as $R=1.25\times R_{\oplus}\left(\frac{M}{M_{\oplus}}\right)^{0.274}$
for a family of planets that have the same Fe/Si ratio. Here, we generalize
the power law relation for low-mass planets to include up to 50\%
H$_{2}$O assuming a fixed ratio between the silicate and metal portion
of a planet. Meaning, that the mantle and core mass fractions are
the same so that the Fe/Si ratio of the family of planets with different
IMF varies little. Figure 3 shows the radius
as a function of mass in terms of Earth-masses for rocky and ocean
planets. The generalization of the power law becomes\[
R_{p}=\left(1+0.56\times\text{IMF}\right)R_{\oplus}\left(\frac{M_{p}}{M_{\oplus}}\right)^{0.262\left(1-0.138\times\text{IMF}\right)}\]
 where IMF denotes the amount of water in percentage. Table 1
shows the values for the coefficient and the exponent in the power
law corresponding to the family of planets with 0 up to 50\% H$_{2}$O.
The coefficient $a$ shows how much larger Earth would be if it had
different amounts of water. Earth would have an increase of 26\% in
radius if it had 50\% water by mass. Our result for a 50\% H$_{2}$O-planet
differ from the exponent obtained by Sotin et. al. possibly because
we include the newly discovered post-perovskite phase in the lower
mantle \citep{Motohiko:2004} and our preference for the Vinet EOS
over the BM3 EOS chosen by Sotin et. al. We are presently working
towards an understanding of this difference.

Figure 3 shows the mass-radius relationship for planets
with a fixed mantle-to-core ratio of 2:1 as a proxy for Fe/Si. It
also shows the relationship between the planetary mass and both the minimum and maximum radii. The minimum radius is obtained for a rocky planet with the
largest possible core constrained from solar nebula composition. The
maximum radius occurs when the planet is made of silicates and water
constrained by the volatiles/Si ratio from chondritic material (refer
to section 4.1).

\section{Observational Characterization of Super-Earth Planets}

The main result from our model calculations presented here is that
there is a threshold of precision in $R$ and $M$ that allows bulk
estimates of the composition $\chi$. These are $\sim$5\% in planetary
radius and $\sim10$\% in planetary mass (this is an average statement
over the mass range $1-10M_{\oplus}$ and our complete grid of models).
Instrumentation currently under development will be able to provide
such observational tolerances in the coming 5 years - in particular,
the NASA $Kepler$ mission in synergy with the HARPS-NEF spectrograph.
 
The $Kepler$ mission \citep{Kepler-Mission} will monitor a single
field in Cygnus/Lyra and discover many hundreds of transiting planets
orbiting preselected main-sequence stars (F to M-type) of 11th to
15th magnitude in V. It is capable of detecting more than 50 Earth-sized
(1$R{}_{\oplus}$) planets in all orbits, with about a dozen n orbits
approaching 1-year period; as well as several hundred super-Earths
($\sim1.3R_{\oplus}$) in all orbits up to 1 year. The $Kepler$ mission
will achieve 20 ppm photometry or better on 12th magnitude stars (its
\char`\"{}sweet spot\char`\"{} for detection). 

The HARPS-NEF spectrograph is a collaborative Harvard-Geneva Observatory
project to build and operate a high-precision radial velocity instrument
in the Northern hemisphere for synergy with $Kepler$. Its design
is based on that of HARPS \citep{Pepe_et_al:2005}; it will be capable of
achieving better than 1m/s in 1 hour on a 12th magnitude star of F-
to M-type. That would allow planet mass determination of 10\% or better
for super-Earths (2-5 $_{\oplus}$) in orbits closer than 0.10 AU.
For more massive super-Earths, 10\% in mass is achieved for orbits
up to 0.5 AU, all for planets already detected with transits by $Kepler$.

Using $Kepler$ and HARPS-NEF together should allow tolerances in
radius ($<5\%$) and mass ($<10\%$) for dozens of super-Earths, limited
only by available observing time for HARPS-NEF. This would produce
a large enough statistical sample to distinguish the bulk properties
of ocean planets and dry rocky planets and their rate of occurance
in orbits closer than $\sim$1AU. Examples of the ability to distinguish
these properties are given in Fig.4 and 5. 

Our estimate of $4\%$ uncertainty in planet radius from $Kepler$ incorporates
the photometric uncertainty for 10 high-impact transits on a 12th magnitude
star and the uncertainty in the stellar parameters \citep{Cody_Sasselov:2002}, and stellar radius in particular. This estimate makes use of the
fact that $Kepler$ will discover a large number of planets and only a
sub-sample of the most suitable ones will be followed-up with HARPS-NEF
and will have $Kepler$-derived parallaxes. A good general discussion of
observational uncertainties for both CoRoT and $Kepler$ is given by
\citet{Selsis_aph:2007}.

The apparent radius of a transiting planet depends on its atmosphere;
\citet{Burrows:2003} have estimated a several percent increase in the
radius of a hot Jupiter compared to a reference radius at 1 bar pressure.
We have neglected this 'transit radius' effect here, but note that it
could be important for super-Earths with significant atmospheres.

\section{Conclusions}
 
We have present here the relationship between planetary mass, radius and bulk
composition for planets in the 1$M_{\oplus}$ to 10$M_{\oplus}$ range. The main
bulk composition materials can be described well in three groups: core (Fe),
mantle (silicates) and water (H$_{2}$O). Therefore the planet radius
dependence on bulk composition for given mass is best illustrated with the
help of a ternary diagram. The degeneracy in composition is illustrated in
curves of constant radius. These iso-radius curves are less sensitive to
mantle composition, showing that the degeneracy is mostly achieved by a
trade-off between core and H$_{2}$O content.

Our detailed models indicate that a maximum terrestrial radius exists for a
given planetary mass - it separates dry rocky compositions from ones
containing $10\%$ or more water by mass. A measured planet radius that exceeds
that critical terrestrial radius would imply that the planet is an ocean
planet. For the high-mass range, the possibility of a massive hydrogen-helium
envelope has to be evaluated separately. 

By considering realistic initial states for our interior models we can
constrain the parameter space of plausible evolved models, as shown in Figs.1,
4, and 5. These constraints lead to minimum and maximum radius as a function
of planet mass, and show that planets with small mantles and large core and
H$_{2}$O contents are unlikely to form. 

We provide a mass-radius relationship to include the effects of various
amounts of water and note that the exponent $\beta$ in
$R=aR_{\oplus}\left(M/M_{\oplus}\right)^{\beta}$ decreases slightly as the
amount of H$_{2}$O increases (due to the high compression of H$_2$O) and $a$
increases almost linearly with H$_{2}$O content (due to the low density of
H$_2$O). 

\section{Acknowledgments}

We thank S. Jacobsen for fruitful discussions and the anonymous reviewer for his/her comments.  
D.V. is grateful for the support from the Harvard Origins of Life Initiative. This work
was supported under NSF grant EAR 04-40017. 

\pagebreak

\begin{deluxetable}{ccc}
\tablewidth{0pt}
\tablecolumns{3}
\tablecaption{Generalized mass-radius relationship -- $R=aR_{\oplus}(\frac{M}{M_{\oplus}})^{\beta}$ }
\tablehead{
\colhead{H$_2$O} & \colhead{a} & \colhead{$\beta$}}
\startdata
 
0 & 1 & 0.262 \\ 
10 & 1.076 & 0.260 \\
20 & 1.130 & 0.257 \\
30 & 1.178 & 0.252 \\
40 & 1.223 & 0.248 \\
50 & 1.265 & 0.244 \\
\enddata
\tablecomments{Values in the power law relationship between mass and radius of a super-Earth with similar Fe/Si ratio to Earth's}
\end{deluxetable}

\begin{figure}
\figurenum{1}
\begin{center}
\includegraphics[scale=0.6]{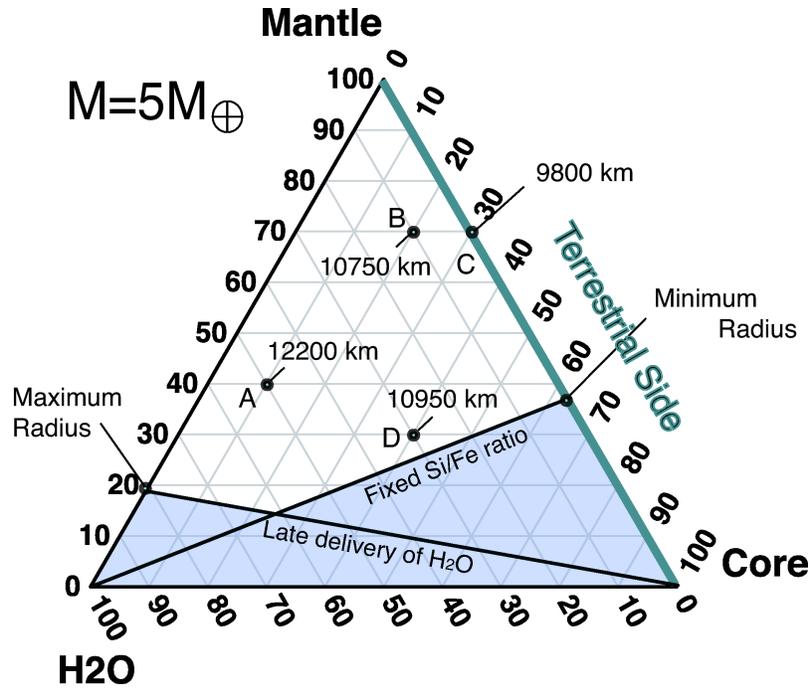}
\end{center}
\caption{
Ternary Diagram. The possible compositions of a super-Earth and radius
are illustrated in a ternary diagram that depicts three axis: amount
of H$_{2}$O, core (Fe) and mantle (silicates). Each point inside
the triangle uniquely specifies a bulk composition and the corresponding
radius - a few examples are labeled. The terrestrial side is the location
where rocky planets would be shown. The minimum plausible radius is achieved
with the largest core possible from solar nebula composition constraints and
no water. The maximum plausible radius is shown for a planet with no core and maximum amount
of water under the solar nebula composition constraints. The shaded region shows improbably compositions from solar
nebula abundances and accretion processes.} 
\end{figure}

\begin{figure}
\figurenum{2}
\begin{center}
\includegraphics[scale=0.6]{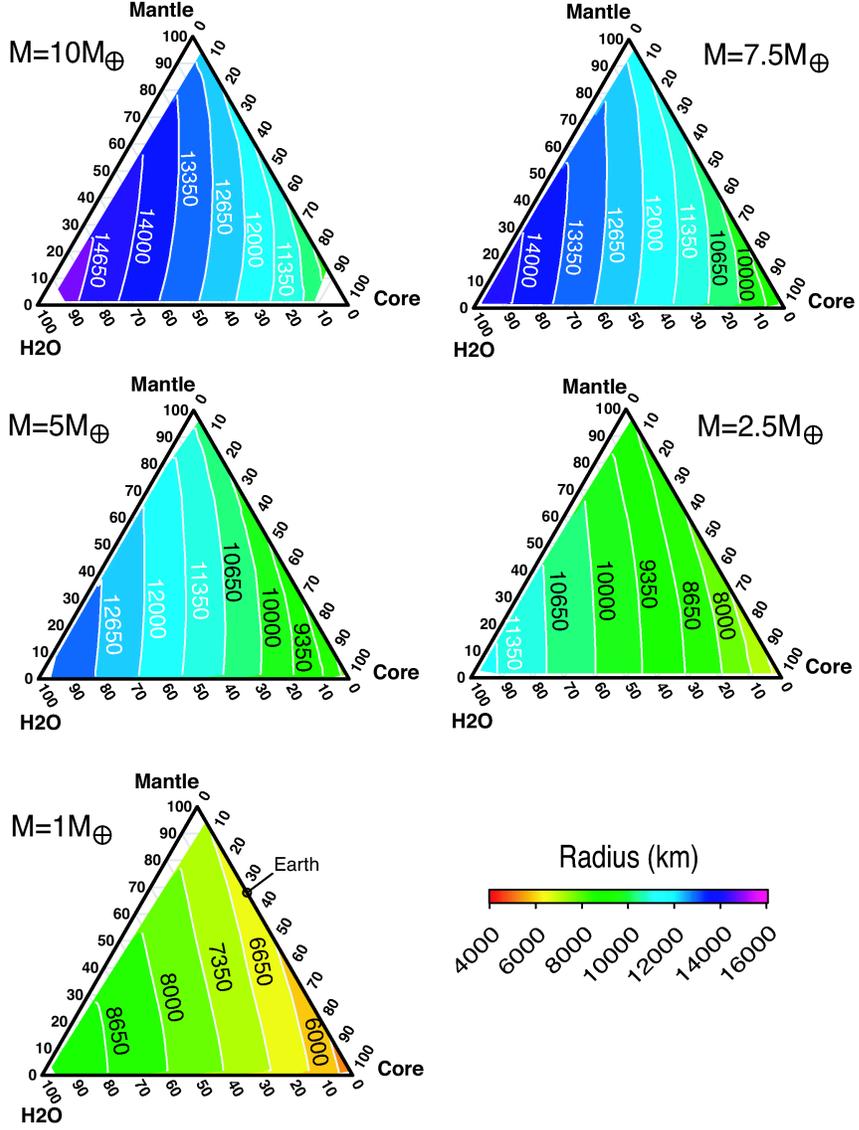}\par\end{center}
\caption{Ternary Diagram relating $M$, $R$,
and $\chi$. The iso-radius curves for possible bulk compositions
are shown for a 10$M_{\oplus}$, 7.5$M_{\oplus}$, 5$M_{\oplus}$, 2.5$M_{\oplus}$ and 1$M_{\oplus}$.
As mass increases, the iso-radius curves shift to the right and rotate
slightly. The progression from a 5000km (1$M_{\oplus}$ Fe-) planet to a 16000km
(10$M_{\oplus}$ water-) planet is illustrated in the color bar. White regions
on the ternary diagrams are areas that pose a numerical challenge
on the model.}
\end{figure}

\begin{figure}
\figurenum{3}
\begin{center}
\includegraphics[scale=0.6]{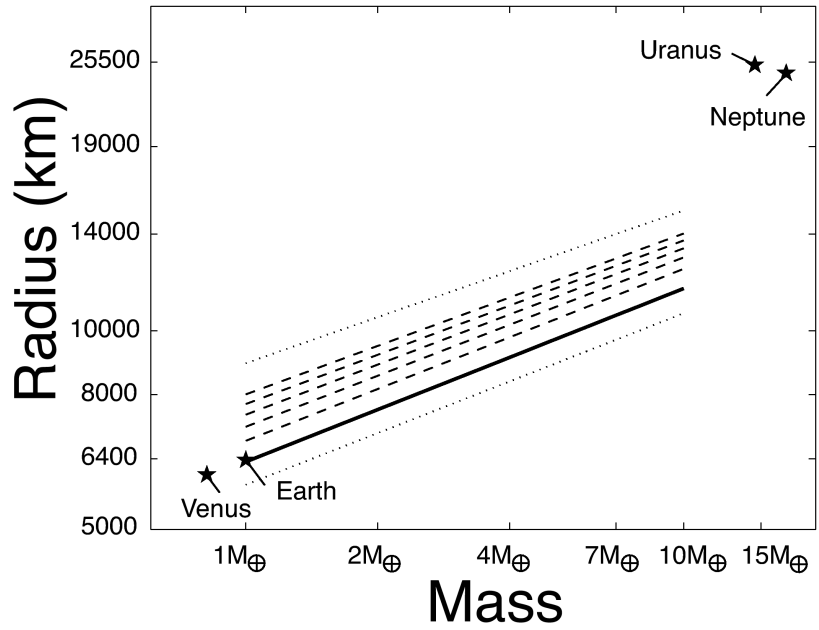}\end{center}

\caption{Mass-Radius relationship for Ocean and
Rocky Massive Planets. The solid black line is the power law relationship
for completely terrestrial planets with 1-10$M_{\oplus}$. The dashed
lines above progressively represent the relationship for planets with
10\%, 20\%, 30\%, 40\% and 50\% H$_{2}$O. This family of planets
have a fixed mantle to core proportion of 2:1. The minimum and maximum
planetary radius relations with mass are shown as dotted lines. Venus,
Earth, Uranus and Neptune are shown for reference.}
\end{figure}
\begin{figure}
\figurenum{4}
\begin{center}
\includegraphics[scale=0.4]{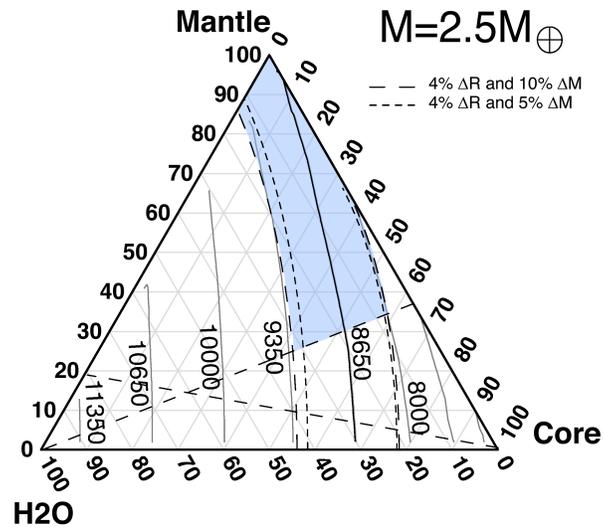}\end{center}

\caption{\emph{Kepler} and HARPS-NEF capabilities for a 2.5 $M_{\oplus}$
planet. The shaded area around the terrestrial radius threshold illustrates
the uncertainty from a 4\% error in radius measurements and 5\% (short-dashed
line) and 10\% (long-dashed line) error mass measurements. The range of possible interior structures (bulk
compositions) is dominated by the radius uncertainty in this range of masses. The two dashed lines define
interior structures excluded by the initial conditions - see Fig.1.
}
\end{figure}

\begin{figure}
\figurenum{5}
\begin{center}\includegraphics[scale=0.6]{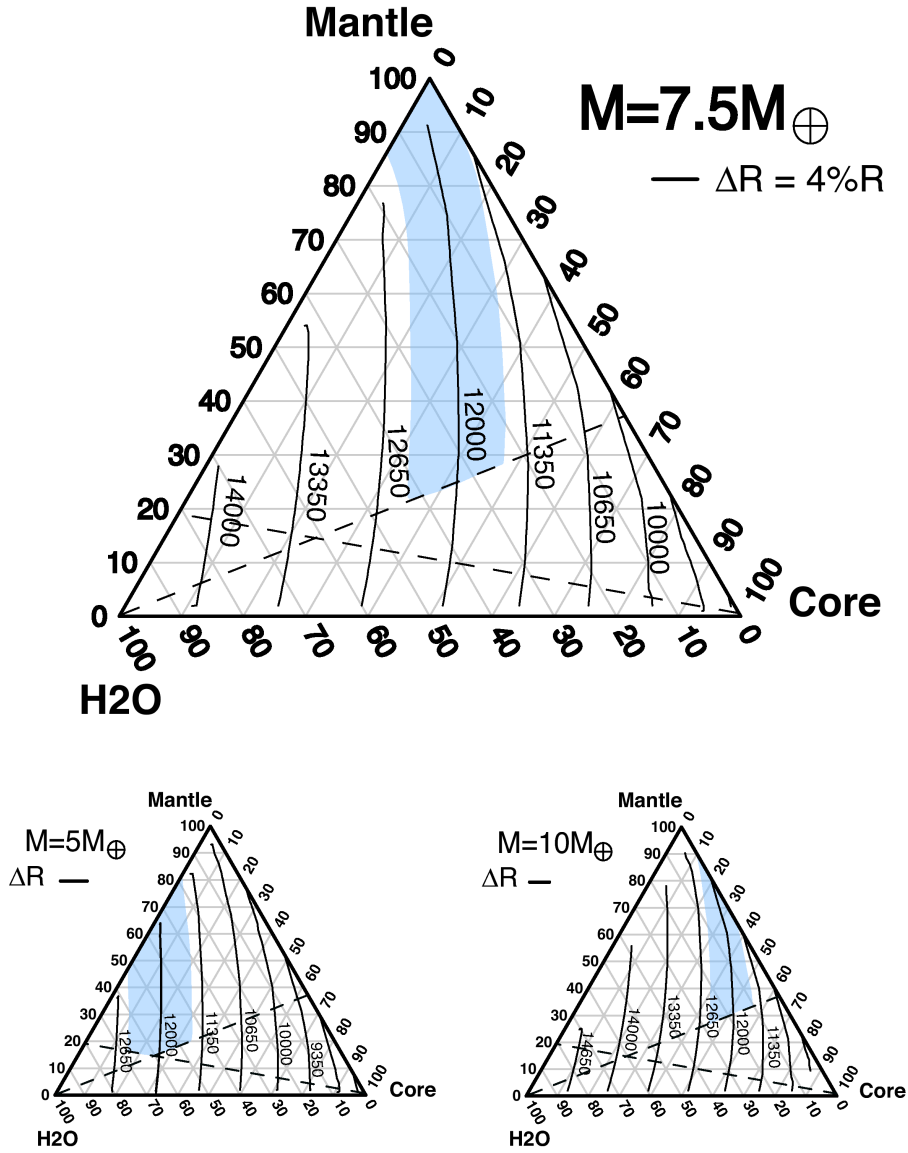}\par\end{center}

\caption{The shaded area around the terrestrial threshold radius illustrates a 4\%
error in radius measurement for a 7.5$M_{\oplus}$ planet. The effect of
uncertainty in planet mass measurement is illustrated in the bottom panel: the same radius value of 12000~km
could correspond to very different compositions, if the planet mass is not
known to better than 30\%. In such a case it is not possible to determine whether the planet is an ocean
planet or a dry rocky planet. The two dashed lines define interior structures excluded by the initial conditions - see Fig.1.}

\end{figure}

\pagebreak
\bibliographystyle{apj}
\bibliography{/home/valencia/Documents/MyPapers/proj}

\end{document}